\documentclass[aps,prx,twocolumn,nofootinbib,longbibliography,superscriptaddress]{revtex4-2}

\usepackage{amsmath}
\usepackage{amssymb}
\usepackage{graphicx}
\usepackage[svgnames]{xcolor}
\usepackage[colorlinks=true, linkcolor=Red , citecolor=NavyBlue, urlcolor=NavyBlue, breaklinks=true]{hyperref}

\renewcommand{\vec}[1]{\boldsymbol{#1}}

\begin{document}

\title{Modified interferometer to measure anyonic braiding statistics}
\author{Steven A.~Kivelson}
\affiliation{Department of Physics, Stanford University, Stanford, CA 94305, USA}
\author{Chaitanya Murthy}
\affiliation{Department of Physics and Astronomy, University of Rochester, Rochester, New York 14627, USA}
\affiliation{Department of Physics, Stanford University, Stanford, CA 94305, USA}

\begin{abstract}
Existing quantum Hall interferometers measure twice the braiding phase, $e^{i2\theta}$, of Abelian anyons, i.e.~the phase accrued when one quasi-particle encircles another clockwise.  
We propose a modified Fabry--P\'{e}rot or Mach--Zehnder interferometer that can measure $e^{i\theta}$.
\end{abstract}

\maketitle

A major advance in the study of topological quantum phases of matter has been achieved through the first direct measurements of the expected fractional braiding statistics of the quasi-particles in the fractional quantum Hall effect using various quantum Hall interferometers~\cite{nakamura2020, nakamura2023, kundu2023, willett2023, mcclure2012, werkmeister2024, samuelson2024, kivelson2020, kivelson2023, feldman2022, chamon1997, camino2005, camino2007, willett2009, ofek2010, halperin2011, ji2003, carrega2021}.
However, as was 
pointed out in Ref.~\cite{read2023}, since these measurements effectively detect the phase accrued as one quasi-particle encircles an integer number of others, they measure twice the statistical phase, $e^{i2\theta}$; one particle encircling another is topologically equivalent to two exchanges of the same handedness.
Among other things, this means that these interferometers could not distinguish bosons ($\theta=0$) from fermions ($\theta=\pi$).  

\begin{figure}
    \centering
    \includegraphics[width=\columnwidth]{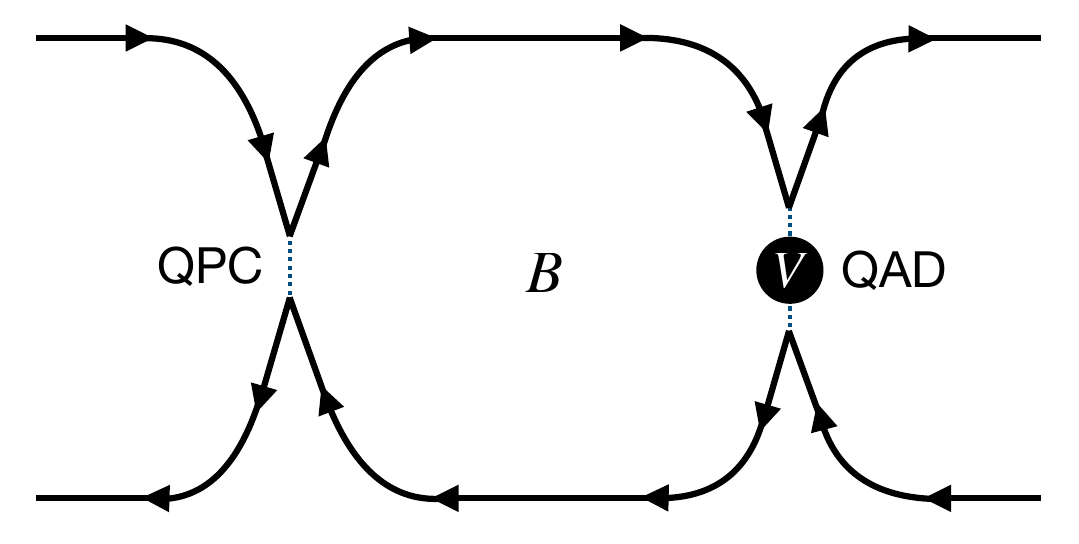}
    \caption{Schematic depiction of the proposed Fabry--P\'{e}rot interferometer with a gate-tunable quantum anti-dot (QAD) midway across one of the two quantum point-contact (QPC) junctions. 
    (Our analysis and conclusions apply equally to a Mach--Zehnder design wherein the other junction is replaced by a continuous edge.)
    The arrows indicate the direction of drift of the chiral edge states. 
    The interferometer is operated in a regime where quasi-particle tunneling across the right junction occurs predominantly via a two-step process through a near-resonant bound state on the QAD, which can be tuned through resonance by varying a gate voltage $V$.
    \label{fig1}}
\end{figure}

Here we propose a slightly more complex version of an interferometer that would directly measure $e^{i\theta}$.%
\footnote{The motivation here is similar to that in an earlier study~\cite{spivak1991, razmadze2020}, in which the sign of the Josephson coupling between two conventional (s-wave) superconductors mediated by tunneling through a quantum dot can be switched from positive to negative depending on whether the dot is unoccupied or singly occupied; the negative Josephson coupling in the latter case derives from the Fermi statistics of the individual electrons that make up the Cooper pair.  
The geometry of the proposed interferometer is similar to one that was explored in a somewhat different context in Refs.~\cite{yacoby1995} and~\cite{schuster1997}.}
The interferometer we have in mind is shown schematically in Fig.~\ref{fig1}.
The basic structure is of a conventional quantum Hall Fabry--P\'{e}rot interferometer, but with a quantum anti-dot (QAD) symmetrically located midway across one of the quantum point contact (QPC) junctions.
(Our discussion applies essentially unchanged to a Mach--Zehnder interferometer~\cite{ji2003} with the QAD midway across its junction).
We assume that the tunneling from the upper (right-moving) edge state to the lower (left-moving) edge state at this junction occurs through a near-resonant two-step process through one of the bound states on the anti-dot.

Specifically, let $\varepsilon$ signify the energy, relative to the chemical potential, of the relevant bound state.
In the ground state of the system, this level is occupied by a quasi-particle if $\varepsilon < 0$, while it is unoccupied if $\varepsilon > 0$.
By tuning a gate voltage, identified as $V$ in the figure, it is possible to tune $\varepsilon$ through zero, i.e.~through the resonance.%
\footnote{The quantization of the states on the anti-dot can be understood through semiclassical quantization of many-anyon bound states, as discussed in Ref.~\cite{kivelson1990} and below.}
When $\varepsilon > 0$, the dominant process simply involves a single quasi-particle tunneling from the upper edge, through the unoccupied near-resonant level, and on to the lower edge. 
When $\varepsilon < 0$, on the other hand, the dominant process is a cooperative one in which the quasi-particle in the occupied near-resonant level first tunnels to the lower edge and then is replaced by a quasi-particle from the upper edge.%
\footnote{Note that these same two processes can equivalently be described in terms of cooperative versus simple resonant tunneling of a quasi-hole from the lower edge to the upper edge, respectively.}

The interferometer is sensitive to the relative phase, $\Theta_{\text{inter}}$ (defined modulo $2\pi$), between processes in which a quasi-particle tunnels through the first point contact and processes in which it tunnels through the second point contact.
This phase can be expressed as
\begin{equation}
\label{eq:phase}
    \Theta_{\text{inter}} 
    = 2\pi \frac{A B}{\phi_0^\star} 
    + 2 \left(N_{\text{qp}}-N_{\text{qh}}\right) \theta 
    + \eta \, \theta \, ,
\end{equation}
where $\phi_0^\star = hc/e^\star$ is the effective magnetic flux quantum with $e^\star$ the factional charge of the quasi-particle, $A$ is the area enclosed in the interferometer, $B$ is the applied perpendicular magnetic field, $N_{\text{qp}}$ and $N_{\text{qh}}$ are, respectively, the number of quasi-particles and quasi-holes enclosed in the interior of the interferometer, and $\theta$ is the statistical phase associated with the clockwise exchange of two quasi-particles.%
\footnote{Note that $2\theta$ and $-2\theta$ are the phases associated with the clockwise transport of a quasi-particle around another quasi-particle and around a quasi-hole, respectively~\cite{arovas1984}.}
The new feature associated with the modified design of the interferometer is the last term, $\eta \, \theta$, where $\eta=0$ for the direct tunneling process through the anti-dot, and $\eta=1$ for the cooperative process. 
The origin of this term can be seen from considering the process in which a quasi-particle begins at the lower edge of the right point contact, propagates along the lower edge and tunnels to the upper edge at the left point contact, propagates back to the right point contact along the upper edge, and finally tunnels back to its original position. 
In the case of direct tunneling through the anti-dot, we have simply taken one quasi-particle around a closed path. 
But in the case of cooperative tunneling, in addition to this we have exchanged the quasi-particle that was originally on the anti-dot with the quasi-particle that was originally at the upper edge of the right point contact.

The idea, then, is to tune the anti-dot through resonance (from $\varepsilon > 0$ to $\varepsilon < 0$) and look for a $\theta$ phase shift in the interference pattern.%
\footnote{The shift of the phase should occur over a range of $\varepsilon$ near $0$ corresponding to the larger of the width of the resonance and the temperature.}
In addition to the fact that this shift is a measure of $\theta$ rather than $2\theta$, an advantage of this design is that the shift is directly induced by a controlled variation of a gate voltage, rather than relying on an accidental (and unknown) rearrangement of quasi-particle occupations inside the interferometer. 
However, a possible confounding factor is that it may be difficult to distinguish a small shift in the enclosed area for the two different resonant processes from the desired statistical shift.%
\footnote{The interference pattern is typically explored by controlling both the magnetic field and the enclosed area by shifting the trajectory of the edge states with a gate.}

\section*{Further considerations}
We now discuss a few subtleties and details that underlie the above analysis:

\vspace{0.8em}
1) As already mentioned, we need to assume that the Aharonov--Bohm (AB) phase accrued is independent of whether the tunneling process through the anti-dot is the direct process or the cooperative one.
It is plausible that this condition is approximately satisfied under various reasonable circumstances, but we have not determined an unambiguous way to insure this.
The details of how the state of the anti-dot might affect the effective area of the interferometer, and hence the AB phase, are complex.
However, the dominant effects can likely be modeled using electrostatic interaction parameters similar to those that determine the change in area due to variations in bulk charge---namely the edge stiffness and bulk-edge coupling~\cite{halperin2011, nakamura2022}.
Thus, if the values of these parameters place a device firmly in the Aharonov--Bohm regime (rather than the Coulomb-dominated regime), one would expect the AB phase accrued to be approximately independent of whether the tunneling process through the anti-dot is the direct process or the cooperative one.
Empirically, in determining whether this condition is satisfied, a first pass should be to look for the expected $\pi$-phase shifts in the integer quantum Hall regime.

\vspace{0.8em}
2) As an aid to intuition, we consider a simple model problem in which a similar process can be used to distinguish bosons from fermions~\cite{konig2002}. 
Consider a system described by the Hamiltonian
\begin{equation}
    H = V \hat{n} 
    + \tfrac{1}{2} U \hat{n} (\hat{n}-1)
    - \sum_{\lambda=u,l}  \left[ t_{\lambda} \hat{c}^\dagger_\lambda \hat{c} + \mathrm{h.c.} \right] ,
\end{equation}
where $\hat{c}_u^\dagger$ and $\hat{c}_l^\dagger$ are either bosonic or fermionic operators that create, respectively, a particle with zero energy on the upper or lower edge, and $\hat{c}^\dagger$ creates a particle on the QAD. 
Here $V$ is the energy of the state on the QAD, $\hat{n} = \hat{c}^\dagger \hat{c}$ is the number of particles on the QAD, and $U>0$ 
is the repulsion between two particles on the QAD.  
(We consider a single spin-polarized level on the QAD, so $U$ is relevant only for the case of bosons.)
To zeroth order in the tunneling (i.e.~when $t_{\lambda} = 0$), the ground state of the QAD has $n(V) \equiv \langle \hat{n} \rangle = 0$ for $V>0$ and $n(V) = 1$ for $-U < V < 0$.
(We will not treat $V < -U$ where, in the bosonic case, $n(V)>1$.)

The effective Hamiltonian that describes the coupling between the edge-states to second order in $t_{\lambda}$ is
\begin{equation}
    H_{\text{eff}} = E_0
    - \sum_{\lambda=u,l} \mu_{\lambda} \hat n_{\lambda} 
    - \left[ \tau \, c_l^\dagger c_u + \mathrm{h.c.} \right] ,
\end{equation}
where, in the fermionic case,
\begin{align}
\label{eq:E0}
    E_0 &= n(V) \left[V+ \sum_\lambda\frac {\lvert t_{\lambda} \rvert^2}{V} \right]
    \, ; \\*
    \tau &= \frac {t_{u}^* t_{l}^{\vphantom{*}}}{V} 
    \ ;  \quad \
    \mu_{\lambda} = \frac {\lvert t_{\lambda} \rvert^2}{V} 
    \, , \notag
\end{align} 
while in the bosonic case, $E_0$ is as in Eq.~(\ref{eq:E0}), and
\begin{align}
    \tau &= t_{u}^* t_{l}^{\vphantom{*}} \left[ \frac{1}{\lvert V \rvert} + \frac{2n(V)}{\lvert U+V \rvert} \right]
    \, ; \\*
    \mu_{\lambda} &= \lvert t_{\lambda} \rvert^2 \left[ \frac{1}{\lvert V \rvert} + \frac{2n(V)}{\lvert U+V \rvert} \right]
     \, . \notag
\end{align} 
In both cases, the magnitude of $\tau$ grows as one approaches the resonance condition, $V=0$;  however, in the bosonic case, the phase of $\tau$ remains the same independent of the sign of $V$, while in the fermionic case there is an abrupt $\pi$ phase shift.%
\footnote{Needless to say, the singular behavior as $|V| \to 0$ is rounded by higher order terms when $|V| \sim |t|$. 
Indeed, in the fermionic version, this is actually a non-interacting problem and readily solved exactly;  in this case, the change in the sign of $\tau$ upon tuning $V$ through $0$ can be understood as the difference between tunneling through an intermediate state with energy greater or less than that of the edge state. 
However, viewed from a many-body perspective, where the Fermi statistics act as an effective interaction, the change in the sign of $\tau$ as $V$ crosses zero is directly related to the fact that for $V>0$, the tunneling process is direct, while for $V<0$ it is an exchange process. 
That this is not simply a question of perspective is established by the comparison with the bosonic case.} 
The overall phase of $\tau$ is of course a matter of convention; it can be altered by a simple redefinition $c_u \to c_u$, $c_l \to e^{i \gamma} c_l$. 
But the abrupt change in the phase of $\tau$ as $V$ changes sign is independent of this ambiguity.

\vspace{0.8em}
3) The spectrum of bound states associated with the anti-dot is ultimately a many-body problem, as interaction effects are always essential to the existence of a fractional quantum Hall state. 
All that is needed for our treatment, however, is a quasi-particle description of the lowest-lying excited states of the anti-dot---either with one added quasi-particle (when $\varepsilon > 0$) or one added quasi-hole (when $\varepsilon < 0$).
The essence of this problem can be understood on the basis of a semiclassical treatment along similar lines as the analysis in Ref.~\cite{kivelson1990}.
We model the QAD as a potential maximum $V(\vec{r})$ that is smooth on the scale of the magnetic length $\ell = \sqrt{\hbar c/eB}$ and weak enough to not mix Landau levels. 
The quasi-hole bound states of the anti-dot, with energies $\epsilon_j < \epsilon_{j+1}$, are associated with semiclassical orbits along equal-potential contours of $V(\vec{r})$. 
The quantization conditions are
\begin{align}
\label{eq:semiclassical_quant}
    2\pi B A_j / \phi_0^\star + 2(j-1) \theta - \pi = 2\pi n_j \, ,
\end{align}
where $A_j$ is the area enclosed by the contour with $V(\vec{r}) = -\epsilon_j$, each $n_j$ is a nonnegative integer, and $j-1$ is the number of quasi-holes enclosed by the trajectory of quasi-hole $j$.
The left hand side of Eq.~(\ref{eq:semiclassical_quant}) is the total phase accumulated by a quasi-hole wavepacket during one full orbit around the contour $V(\vec{r}) = -\epsilon_j$.
The first term is the Aharonov-Bohm phase corresponding to the enclosed magnetic flux, while the second term is the statistical phase associated with braiding around the $j-1$ enclosed quasi-holes.
The extra $-\pi$ is the ``Maslov correction'' accounting for additional phase shifts of the wavepacket as it passes through classical turning points of the orbit.%
\footnote{The same Maslov correction appears in the semiclassical treatment of the one-dimensional harmonic oscillator, where it leads to the familiar zero-point energy shift in the spectrum.}
Bound states occur when the total phase accumulated is an integer multiple of $2\pi$.
A many-body state of the QAD is then specified by a set of occupied orbitals that satisfy Eq.~(\ref{eq:semiclassical_quant}) with some sequence of integers, $n_j$.

The ground state with $M$ quasi-holes, $\psi_M$, is obtained by occupying the first $M$ orbitals which satisfy Eq.~(\ref{eq:semiclassical_quant}), $\epsilon_1, \epsilon_2, \dots, \epsilon_M$, with all $n_j = 0$.
We take the near-resonant level to be the $M$th orbital, and $\varepsilon = -\epsilon_M - \mu$.
Thus, when $\varepsilon > 0$, the ground state of the QAD is $\psi_M$ and the first excited many-body state is $\psi_{M-1}$ (the lowest-energy state with one less quasi-hole, i.e.~one added quasi-particle).
When $\varepsilon < 0$, the ground state is instead $\psi_{M-1}$ and the first excited state is $\psi_M$.%
\footnote{In going from the $\varepsilon > 0$ ground state $\psi_M$ to the $\varepsilon < 0$ ground state $\psi_{M-1}$, we assume that the quasi-hole removed from the QAD is trapped in some distant region of the sample, outside the interferometer.}

\begin{figure}
    \centering
    \includegraphics[width=\columnwidth]{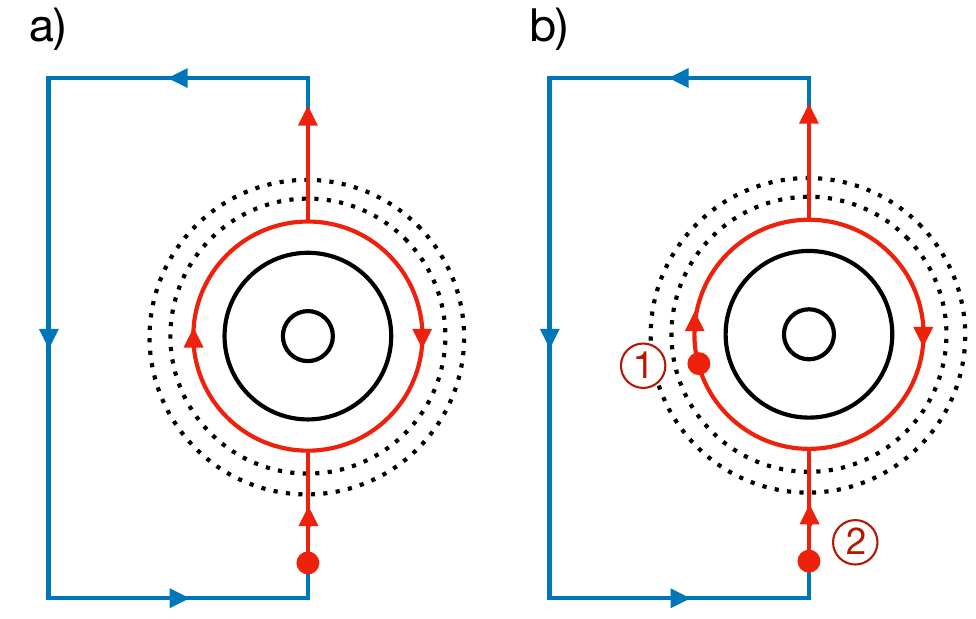}
    \caption{Schematic for the semiclassical treatment of quasi-hole tunneling through the QAD.
    Trajectories that contribute to direct and cooperative tunneling through the near-resonant quasi-hole bound state on the QAD are shown in panels a and b, respectively.
    In order to assign definite phases to each process, we include identical circuits of the tunneled quasi-hole around the body of the interferometer so as to return to the starting configurations.
    (The arrows are reversed relative to Fig.~\ref{fig1} because we are considering quasi-hole rather than quasi-particle motion here.)
    Solid/dotted black circles represent semiclassical orbits corresponding to other occupied/unoccupied quasi-hole bound states on the QAD.
    In panel b depicting cooperative tunneling, the two quasi-holes involved are labelled ``1'' and ``2'', respectively.
    \label{fig2}}
\end{figure}

Now consider the near-resonant direct tunneling of a quasi-hole through the QAD when $\varepsilon < 0$.
Semiclassically, this corresponds to a coherent sum over trajectories in which the quasi-hole arrives with energy $\epsilon \approx \epsilon_M$, orbits $m+\frac{1}{2}$ times around the contour $V(\vec{r}) = -\epsilon$, and departs, as depicted in Fig.~\ref{fig2}a.
We can assign a definite phase to each trajectory if we append a circuit of the quasi-hole around the body of the interferometer to return it to its starting position.
The total phase factor for a trajectory includes both the Aharonov-Bohm phase corresponding to the enclosed magnetic flux and the statistical phase associated with anyon braiding ($\theta$ for each clockwise exchange of quasi-holes, or $2\theta$ each time a quasi-hole encircles another clockwise).
The resonance condition, that the trajectories with different $m$ interfere constructively, is the same as the quantization condition~(\ref{eq:semiclassical_quant}), which ensures that the total phase accrued during each full orbit around the QAD is a multiple of $2\pi$.
Similarly, the near-resonant cooperative tunneling process of a quasi-hole when $\varepsilon > 0$ corresponds to a coherent sum over trajectories in which a quasi-hole originally on the QAD with energy $\epsilon \approx \epsilon_M$ orbits $m' + \frac{1}{2}$ times around the contour $V(\vec{r}) = -\epsilon$ and departs, after which another quasi-hole with energy $\epsilon$ arrives and orbits $m''$ times around the same contour, as depicted in Fig.~\ref{fig2}b
(the two quasi-holes involved are labelled ``1'' and ``2'', respectively).

To compute the relative phase between the direct tunneling process and the cooperative one, we note that in either case, as discussed above, the total phase accrued during a full orbit around the anti-dot is a multiple of $2\pi$. 
Therefore it suffices to compare the direct trajectory with $m=0$ to the cooperative trajectory with $m' = m'' = 0$.
These differ in that the latter involves a counter-clockwise exchange of quasi-holes, whereas the former does not, and hence the relative phase is $-\theta$.%
\footnote{Strictly speaking, we should compare trajectories in which the quasi-holes start and end at the same locations to unambiguously determine the relative phase. 
This is not quite the case in the discussion above because, in the direct trajectory $T$, the second quasi-hole is located in a distant region of the sample (not shown in the figure). 
However, it suffices to consider a modified trajectory $T'$ in which this quasi-hole starts on the QAD, adiabatically moves to the distant location along some path $P$, and then, after the direct tunneling process has concluded, returns to the QAD along the inverse path $P^{-1}$. 
By construction, the trajectories $T$ and $T' = P^{-1}TP$ carry the same phase, and the latter can be directly compared to the cooperative trajectory.}
Recalling that the direct quasi-hole tunneling process corresponds to the cooperative quasi-particle process, and vice versa, we recover the $\eta \, \theta$ term in Eq.~(\ref{eq:phase}).

\vspace{0.8em}
4) We have also analyzed the problem in terms of the Berry phase accrued for various assumed patterns of adiabatic motion of quasi-holes using the wave-function approach of Arovas, Schrieffer and Wilczek~\cite{arovas1984,stone1992}. 
The calculations are outlined in the Appendix. 
We are able to reproduce the results of the semiclassical analysis described above.
However, not surprisingly, the result depends in detail on the precise location of the assumed trajectory followed, and so does not obviously help to resolve the important issue of disentangling the statistical from the geometric contributions to the phase.

\vspace{0.8em}
5) While the statistical interaction between quasi-particles is the longest-range interaction, other shorter-range interactions manifestly exist as well.  
These can result in geometric differences in the paths followed by quasi-particles undergoing direct versus cooperative tunneling processes, which in turn can result in differences in the Berry phase that have nothing to do with braiding statistics.  
For this reason, we do not necessarily expect the present measurement to be as precise as the measurement of $2\theta$ associated with braiding about a quasi-particle in the bulk, as the latter can in principle be spatially far separated from the propagating modes of the interferometer.  
However, given an accurate measurement of $2\theta$, the present interferometer need only be accurate enough to distinguish $\theta$ from $\theta + \pi$.

We do not know of any simple way to estimate how large such errors might be in any given experiment---a proper estimate would require a numerical solution of the interacting many-body problem associated with a realistic QAD potential, which is well beyond the scope of our work.
However, evidence for the smallness of such errors can be sought empirically in various ways.
First, as discussed in point 1 above, one should look for the expected $\pi$-phase shifts in the integer quantum Hall regime.
Second, one should check that the same phase shift $\theta$ occurs across multiple resonances of the QAD in the fractional regime.
It seems unlikely that short-range interaction effects would lead to systematic phase shifts of this form.
Finally, the obtained value of $\theta$ should be compared with the value of $2\theta$ extracted from random phase slips, in the same device, associated with braiding about quasi-particles in the bulk.
Since, as discussed above, the latter are less sensitive to short-range interaction effects, consistency between these values would suggest that the relevant errors are small.

\vspace{0.8em}
6) The above analysis applies directly to simple ``Laughlin'' states with Abelian anyons and (under appropriate conditions, i.e.~in the absence of any non-trivial form of edge reconstruction) a single edge mode. 
The extension of these considerations to hierarchical states~\cite{jain1993} and most importantly to states with non-Abelian anyons~\cite{stern2006, bonderson2006, feldman2006} is presumably possible.

\begin{acknowledgments}
We acknowledge helpful comments from Moty Heiblum, Patrick Lee, Hans Hansson, Steven Simon, Claudio Chamon, Eduardo Fradkin, and Nicholas Read, and inspiring discussions with James Ehrets, Thomas Werkmeister, Christina Henzinger, Philip Kim, Benjamin Sac\'{e}p\'{e}, and Bernd Rosenow.
This work was supported in part by the Department of Energy, Office of Basic Energy Sciences, under contract no.~DE AC02-76SF00515 (SAK and CM), and in part by the Gordon and Betty Moore Foundation's EPiQS Initiative through GBMF8686, as well as startup funds from the University of Rochester (CM).
\end{acknowledgments}

\clearpage

\clearpage
\appendix

\section{The wavefunction approach}

Here, we outline the Berry phase calculation of quasi-hole tunneling through the QAD using the wavefunction approach of Arovas, Schrieffer and Wilczek~\cite{arovas1984,stone1992}.

The un-normalized wavefunction describing $M$ quasi-holes at positions $u_a \in \mathbb{C}$ in a $\nu = 1/m$ Laughlin state on the plane, in symmetric gauge, is
\begin{align}
\psi(\vec{z};\vec{u}) = \prod_{i,a} (z_i - u_a) \prod_{i < j} (z_i - z_j)^m \, e^{- \sum_k |z_k|^2 / 4 \ell^2} .
\end{align}
The holomorphic and anti-holomorphic components of the Berry connection over the configuration space of the quasi-holes are given by
\begin{align}
A_a = - \frac{i}{2} \frac{\partial \log Z}{\partial u_a} , \qquad
A_{\bar{a}} = \frac{i}{2} \frac{\partial \log Z}{\partial \bar{u}_a} ,
\end{align}
where $Z = \langle \psi(\vec{u}) \vert \psi(\vec{u}) \rangle$ is the norm of the wavefunction.
Using the plasma analogy to compute $Z$ yields
\begin{align}
A_a &= - \frac{i}{2m} \sum_{b \neq a} \frac{1}{u_a - u_b} + \frac{i \bar{u}_a}{4m \ell^2} , \\*
A_{\bar{a}} &= \, \frac{i}{2m} \sum_{b \neq a} \frac{1}{\bar{u}_a - \bar{u}_b} - \frac{i u_a}{4m \ell^2} ,
\end{align}
which hold as long as the quasi-holes are far from one another, relative to the screening length $\lambda$ of the plasma.
The Berry phase accrued as the quasi-holes traverse any closed path $C$ in their configuration space is then
\begin{align}
\gamma = - \oint_C (A_a du_a  + A_{\bar{a}} du_{\bar{a}}) .
\end{align}

\begin{figure}
    \centering
    \includegraphics[width=\columnwidth]{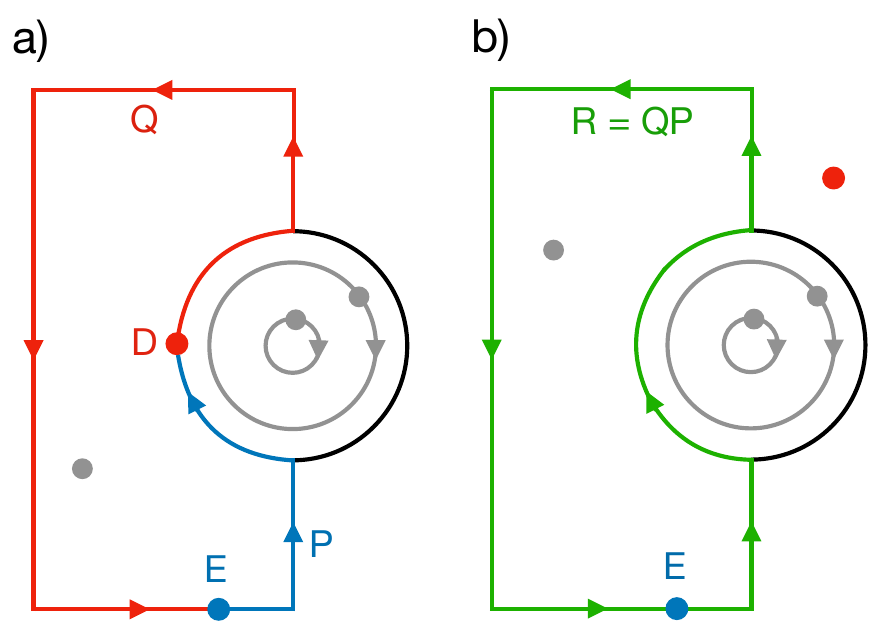}
    \caption{
    Schematic trajectories for the Berry phase calculation of quasi-hole tunneling through the QAD.
    Trajectories for cooperative and direct tunneling through the near-resonant level on the QAD are shown in panels a and b, respectively.
    In the cooperative process (panel a), the two participating quasi-holes move along the blue path $P$ from $E$ to $D$ and the red path $Q$ from $D$ to $E$, respectively.
    In the direct process (panel b), the tunneling quasi-hole moves along the green loop $R = QP$ from $E$ back to $E$; the quasi-hole that occupied the near-resonant level in the cooperative case is now localized outside the interferometer loop (red dot).
    Various other ``spectator'' quasi-holes, and their assumed trajectories, are shown in gray.
    \label{fig3}}
\end{figure}

To model the cooperative quasi-hole tunneling process when the near-resonant anti-dot level is occupied by a quasi-hole, we consider two quasi-holes initially located at a point $E$ on the edge and a point $D$ on the anti-dot, respectively.
The first quasi-hole adiabatically moves along a path $P$ from $E$ to $D$ while the second simultaneously moves along a path $Q$ from $D$ to $E$, as shown schematically in Fig.~\ref{fig3}a.
Since the quasi-holes are identical, this constitutes a closed loop in their configuration space.
The associated Berry phase, $\gamma_1$, is calculated using the above formulae.
Note that the moving quasi-holes never approach each other closely during this process.

To model direct quasi-hole tunneling when the near-resonant level is empty (i.e.~occupied by a quasi-particle), we consider a single quasi-hole initially located at $E$, which then adiabatically moves along the closed loop $R = Q P$, as shown in Fig.~\ref{fig3}b.
The associated Berry phase, $\gamma_2$, is calculated as above.
The difference in Berry phases for the two processes is easily found to be
\begin{align}
\gamma_1 - \gamma_2 = \frac{\pi}{m} = \theta .
\end{align}
This result is independent of the locations of any other stationary quasi-holes, as long as the total number of these inside the contour $R = Q P$ is the same in the two processes considered.

If the trajectories followed by the moving quasi-holes are slightly different in the two processes, $R \neq QP$, then one finds that $\gamma_1 - \gamma_2$ deviates from $\theta$ by an amount equal to the Aharonov--Bohm phase associated with the closed contour $R^{-1} QP$.
Similarly, if one allows other ``spectator'' quasi-holes to move, the result deviates from $\theta$ if the motion of these spectators differs in the two processes considered.

Thus, while the wavefunction approach does give some insight into the possible sources of error in the tunneling phase shift across a resonance, it does not obviously help in quantifying these errors, since they depend on the assumed quasi-hole trajectories for the direct and cooperative processes.
In addition, the validity of the approximations involved is unclear for the essential part of the trajectory in which the quasi-hole(s) move through the anti-dot region, since at this point they would, in reality, necessarily pass close to other quasi-holes bound to the anti-dot.
For these reasons, the wavefunction approach should not be regarded as superseding the semiclassical analysis presented in point 3 of the main text.

\clearpage

\end{document}